\documentclass{aa}

\usepackage{graphicx}
\usepackage{txfonts}

\usepackage{xcolor}

\usepackage[pdftitle={An alternative interpretation of the exomoon candidate signal in the combined  Kepler and Hubble data of Kepler-1625}, colorlinks = true, breaklinks = true, citecolor = blue, linkcolor = blue, urlcolor = blue, pdfauthor = {Ren\'{e} Heller}]{hyperref}

\usepackage{esvect}  

\begin{document}

\title{An alternative interpretation of the exomoon candidate signal\\in the combined {\it Kepler} and {\it Hubble} data of Kepler-1625}


\titlerunning{Alternative explanations for the exomoon interpretation for Kepler-1625}

\author{Ren\'{e} Heller\inst{1}
          \and
          Kai Rodenbeck\inst{2,1}
          \and
          Giovanni Bruno\inst{3}
          }

   \institute{Max Planck Institute for Solar System Research, Justus-von-Liebig-Weg 3, 37077 G\"ottingen, Germany\\ \href{mailto:heller@mps.mpg.de}{heller@mps.mpg.de}, \href{mailto:rodenbeck@mps.mpg.de}{rodenbeck@mps.mpg.de}
   \and
   Institute for Astrophysics, Georg August University G{\"o}ttingen, Friedrich-Hund-Platz 1, 37077 G{\"o}ttingen, Germany
   \and
   INAF, Astrophysical Observatory of Catania, Via S. Sofia 78, 95123 Catania, Italy, \href{mailto:giovanni.bruno@inaf.it}{giovanni.bruno@inaf.it}
             }

   \date{Received 18 December 2018; Accepted 25 February 2019}

 
  \abstract
{{\it Kepler} and {\it Hubble} photometry of a total of four transits by the Jupiter-sized exoplanet Kepler-1625\,b have recently been interpreted to show evidence of a Neptune-sized exomoon. The key arguments were an apparent drop in stellar brightness after the planet's October 2017 transit seen with {\it Hubble} and its 77.8\,min early arrival compared to a strictly periodic orbit.}
{The profound implications of this first possible exomoon detection and the physical oddity of the proposed moon, i.e., its giant radius prompt us to examine the planet-only hypothesis for the data and to investigate the reliability of the Bayesian information criterion (BIC) used for detection.}
{We combined {\it Kepler}'s Pre-search Data Conditioning Simple Aperture Photometry (PDCSAP) with the previously published {\it Hubble} light curve. In an alternative approach, we performed a synchronous polynomial detrending and fitting of the {\it Kepler} data combined with our own extraction of the {\it Hubble} photometry. We generated five million parallel-tempering Markov chain Monte Carlo (PTMCMC) realizations of the data with both a planet-only model and a planet-moon model, and compute the BIC difference ($\Delta$BIC) between the most likely models, respectively.}
{The $\Delta$BIC values of $-44.5$ (using previously published {\it Hubble} data) and $-31.0$ (using our own detrending) yield strong statistical evidence in favor of an exomoon. Most of our orbital realizations, however, are very different from the best-fit solutions, suggesting that the likelihood function that best describes the data is non-Gaussian. We measure a 73.7\,min early arrival of Kepler-1625\,b for its {\it Hubble} transit at the $3\,\sigma$ level. This deviation could be caused by a 1\,d data gap near the first {\it Kepler} transit, stellar activity, or unknown systematics, all of which affect the detrending. The radial velocity amplitude of a possible unseen hot Jupiter causing the Kepler-1625\,b transit timing variation could be approximately 100\,m\,s$^{-1}$.}
{Although we find a similar solution to the planet-moon model to that previously proposed, careful consideration of its statistical evidence leads us to believe that this is not a secure exomoon detection. Unknown systematic errors in the {\it Kepler}/{\it Hubble} data make the $\Delta$BIC an unreliable metric for an exomoon search around Kepler-1625\,b, allowing for alternative interpretations of the signal.}


   \maketitle
%

\section{Introduction}
\label{sec:introduction}

The recent discovery of an exomoon candidate around the transiting Jupiter-sized object Kepler-1625\,b orbiting a slightly evolved solar mass star \citep{2018AJ....155...36T} came as a surprise to the exoplanet community. This Neptune-sized exomoon, if confirmed, would be unlike any moon in the solar system, it would have an estimated mass that exceeds the total mass of all moons and rocky planets of the solar system combined. It is currently unclear how such a giant moon could have formed \citep{2018A&A...610A..39H}.

\citet{2018A&A...617A..49R} revisited the three transits obtained with the {\it Kepler} space telescope between 2009 and 2013 and found marginal statistical evidence for the proposed exomoon. Their transit injection-retrieval tests into the out-of-transit {\it Kepler} data of the host star also suggested that the exomoon could well be a false positive. A solution to the exomoon question was supposed to arrive with the new {\it Hubble} data of an October 2017 transit of Kepler-1625\,b \citep{2018SciA....4.1784T}.

The new evidence for the large exomoon by \citet{2018SciA....4.1784T}, however, remains controversial. On the one hand, the {\it Hubble} transit light curve indeed shows a significant decrease in stellar brightness that can be attributed to the previously suggested moon. Perhaps more importantly, the transit of Kepler-1625\,b occurred 77.8\,min earlier than expected from a sequence of strictly periodic transits, which is in very good agreement with the proposed transit of the exomoon candidate, which occurred before the planetary transit. On the other hand, an upgrade of {\it Kepler}'s Science Operations Center pipeline from version 9.0 to version 9.3 caused the exomoon signal that was presented in the Simple Aperture Photometry (SAP) measurements in the discovery paper \citep{2018AJ....155...36T} to essentially vanish in the SAP flux used in the new study of \citet{2018SciA....4.1784T}. This inconsistency, combined with the findings of \citet{2018A&A...617A..49R} that demonstrate that the characterization and statistical evidence for this exomoon candidate depend strongly on the methods used for data detrending, led us to revisit the exomoon interpretation in light of the new {\it Hubble} data.

Here we address two questions. How unique is the proposed orbital solution of the planet-moon system derived with the Bayesian information criterion (BIC)? What could be the reason for the observed 77.8\,min difference in the planetary transit timing other than an exomoon?

\section{Methods}
\label{sec:methods}

Our first goal was to fit the combined {\it Kepler} and {\it Hubble} data with our planet-moon transit model \citep{2018A&A...617A..49R} and to derive the statistical likelihood for the data to represent the model. In brief, we first model the orbital dynamics of the star-planet-moon system using a nested approach, in which the planet-moon orbit is Keplerian and unperturbed by the stellar gravity. The transit model consists of two black circles, one for the planet and one for the moon, that pass in front of the limb-darkened stellar disk. The resulting variations in the stellar brightness are computed using Ian Crossfield's {\tt python} code of the \citet{2002ApJ...580L.171M} analytic transit model.\footnote{Available at \href{http://www.astro.ucla.edu/~ianc/files}{www.astro.ucla.edu/${\sim}$ianc/files}} The entire model contains 16 free parameters and it features three major updates compared to \citet{2018A&A...617A..49R}: (1) Planet-moon occultations are now correctly simulated, (2) the planet's motion around the local planet-moon barycenter is taken into account, and (3) inclinations between the circumstellar orbit of the planet-moon barycenter and the planet-moon orbit are now included.

We used the {\tt emcee} code\footnote{Available at \href{http://dfm.io/emcee/current/user/pt}{http://dfm.io/emcee/current/user/pt}} of \citet{2013PASP..125..306F} to generate Markov chain Monte Carlo (MCMC) realizations of our planet-only model ($\mathcal{M}_0$) and planet-moon model ($\mathcal{M}_1$) and to derive posterior probability distributions of the set of model parameters ($\vv{\theta}$). We tested both a standard MCMC sampling with 100 walkers and a parallel-tempering ensemble MCMC (PTMCMC) with five temperatures, each of which has 100 walkers. As we find a better convergence rate for the PTMCMC sampling, we use it in the following. Moreover, PTMCMC can sample both the parameter space at large and in regions with tight peaks of the likelihood function. The PTMCMC sampling is allowed to walk five million steps.

The resulting model light curves are referred to as $\mathcal{F}_{i}(t,\vv{\theta})$, where $t$ are the time stamps of the data points from {\it Kepler} and {\it Hubble} ($N$ measurements in total), for which time-uncorrelated standard deviations $\sigma_j$ at times $t_j$ are assumed, following the suggestion of \citet{2018SciA....4.1784T}. This simplifies the joint probability density of the observed (and detrended) flux measurements ($F(t)$) to the product of the individual probabilities for each data point,

\begin{align}
p(F|\vv{\theta},\mathcal{M}_i)=\prod_{j=1}^{N} \frac{1}{\sqrt{2\pi\sigma_j^2}}\exp\left(-\frac{\left(F(t_j)-\mathcal{F}_i (t_j,\vv{\theta})\right)^2}{2\sigma_j^2}\right) \ .
\end{align}

\noindent
We then determined the set of parameters ($\vv{\theta}_{\rm max}$) that maximizes the joint probability density function ($p(F|\vv{\theta}_{\rm max},\mathcal{M}_i)$) for a given light curve $F(t_j)$ and model $\mathcal{M}_i$ and calculated the BIC \citep{1978AnSta...6..461S}

\begin{align}\label{eq:BIC}
 \text{BIC}(\mathcal{M}_i|F)=m_{i}\ln N-2\ln p(F|\vv{\theta}_{\rm max},\mathcal{M}_i) \ .
\end{align}

\noindent

\noindent
The advantage of the BIC in comparison to $\chi^2$ minimization, for example, is in its relation to the number of model parameters ($m_{i}$) and data points. The more free parameters in the model, the stronger the weight of the first penalty term in Eq.~\eqref{eq:BIC}, thereby mitigating the effects of overfitting. Details of the actual computer code implementation or transit simulations aside, this Bayesian framework is essentially what the {\it Hunt for Exomoons with Kepler} survey used to identify and rank exomoon candidates \citep{2012ApJ...750..115K}, which ultimately led to the detection of the exomoon candidate around Kepler-1625\,b after its first detection via the orbital sampling effect \citep{2014ApJ...787...14H,2016ApJ...820...88H}.

\subsection{Data preparation}

In a first step, we used {\it Kepler}'s Pre-search Data Conditioning Simple Aperture Photometry (PDCSAP) and the {\it Hubble} Wide Field Camera 3 (WFC3) light curve as published by \citet{2018SciA....4.1784T} based on their quadratic detrending. Then we executed our PTMCMC fitting and derived the $\Delta$BIC values and the posterior parameter distributions.


In a second step, we did our own extraction of the {\it Hubble} light curve including an exponential ramp correction for each {\it Hubble} orbit. Then we performed the systematic trend correction together with the transit fit of a planet-moon model. Our own detrending of the light curves is not a separate step, but it is integral to the fitting procedure. For each calculation of the likelihood, we find the best fitting detrending curve by dividing the observed light curve by the transit model and by fitting a third-order polynomial to the resulting light curve. Then we remove the trend from the original light curve by dividing it through the best-fit detrending polynomial and evaluate the likelihood. We also performed a test in which the detrending parameters were free PTMCMC model parameters and found similar results for the parameter distributions but at a much higher computational cost. We note that the resulting maximum likelihood is (and must be) the same by definition if the PTMCMC sampling converges.

Kepler-1625 was observed by {\it Hubble} under the GO program 15149 (PI Teachey). The observations were secured from October 28 to 29, 2017, to cover the $\sim20$\,hr transit plus several hours of out-of-transit stellar flux \citep{2018SciA....4.1784T}. The F130N filter of WFC3 was used to obtain a single direct image of the target, while 232 spectra were acquired with the G141 grism spanning a wavelength range from 1.1 to $1.7\,\mu$m. Due to the faintness of the target, it was observed in staring mode \citep[e.g.][]{berta2012,wilkins2014} unlike the most recent observations of brighter exoplanet host stars, which were monitored in spatial scanning mode \citep{mccullough2012}. Hence, instead of using the IMA files as an intermediate product, we analyzed the FLT files, which are the final output of the \texttt{calwfc3} pipeline of \textit{Hubble} and allow a finer manipulation of the exposures during consecutive nondestructive reads. Each FLT file contains measurements between about 100 and 300 electrons per second, with exposure times of about 291 seconds.

\begin{figure}[t!]
\centering
\includegraphics[width=0.36\textwidth]{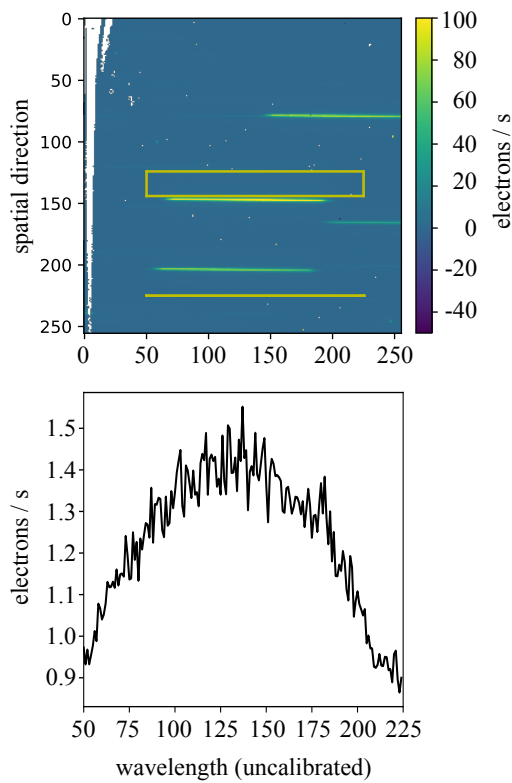}
\caption{{\it Top}: Example of a WFC3 exposure of Kepler-1625. The abscissa shows the column pixel prior to wavelength calibration. The yellow box indicates the region used for background estimation. The spectrum of Kepler-1625 is at the center of the frame, around row 150 in the spatial direction, while several contaminant sources are evident in other regions of the detector. The color bar illustrates the measured charge values. {\it Bottom}: Background value measured across the rows of the same frame.}
\label{fig:HST}
\end{figure}

We used the centroid of the stellar image to calculate the wavelength calibration, adopting the relations of \cite{pirzkal2016}. For each spectroscopic frame, we first rejected the pixels flagged by \texttt{calwfc3} as ``bad detector pixels'', pixels with unstable response, and those with uncertain flux value (Data Quality condition 4, 32, or 512). Then we corrected each frame with the flat field file available on the Space Telescope Science Institute (STScI) website\footnote{\href{www.stsci.edu/hst/wfc3/analysis/grism\_obs/calibrations/wfc3\_g141.html}{www.stsci.edu/hst/wfc3/analysis/grism\_obs/calibrations/}\\\href{www.stsci.edu/hst/wfc3/analysis/grism\_obs/calibrations/wfc3\_g141.html}{wfc3\_g141.html}} by following the prescription of the WFC3 online manual. 
We performed the background subtraction on a column-by-column basis. Due to a number of contaminant stars in the observation field (Fig.~\ref{fig:HST}, top panel), we carefully selected a region on the detector that was as close as possible to the spectrum of Kepler-1625, close to row 150 in spatial dimension, and far from any contaminant. For each column on the detector, we applied a $5\,\sigma$ clipping to reject the outliers and then calculated the median background flux value in that column. Following STScI prescriptions, we also removed pixels with an electron-per-second count larger than 5. An example for the background behavior is shown in the bottom panel of Fig.~\ref{fig:HST}.

We inspected each frame with the {\tt image registration} package \citep{baker2001} to search drifts in both axes of the detector with respect to the very last frame, and then extracted the spectrum of Kepler-1625 by performing optimal extraction \citep{horne1986} on the detector rows containing the stellar flux. This procedure automatically removes bad pixels and cosmic rays from the frames by correcting them with a smoothing function. We started the extraction with an aperture of a few pixels centered on the peak of the stellar trace and gradually increased its extension by one pixel per side on the spatial direction until the flux dispersion reached a minimum.

We performed another outlier rejection by stacking all the one-dimensional spectra along the time axis. We computed a median-filtered version of the stellar flux at each wavelength bin and performed a $3\,\sigma$ clipping between the computed flux and the median filter. Finally, we summed the stellar flux across all wavelength bins from 1.115 to 1.645 $\mu$m to obtain the band-integrated stellar flux corresponding to each exposure.

Before performing the PTMCMC optimization, we removed the first {\it Hubble} orbit from the data set and the first data point of each {\it Hubble} orbit, as they are affected by stronger instrumental effects than the other observations \citep{deming2013} and cannot be corrected with the same systematics model. We also removed the last point of the 12th, 13th, and 14th \textit{Hubble} orbit since they were affected by the passage of the South Atlantic Anomaly (as highlighted in the proposal file, available on the STScI website).

\subsection{Proposed unseen planet}

\begin{figure*}[t!]
\centering
\includegraphics[width=0.49\textwidth]{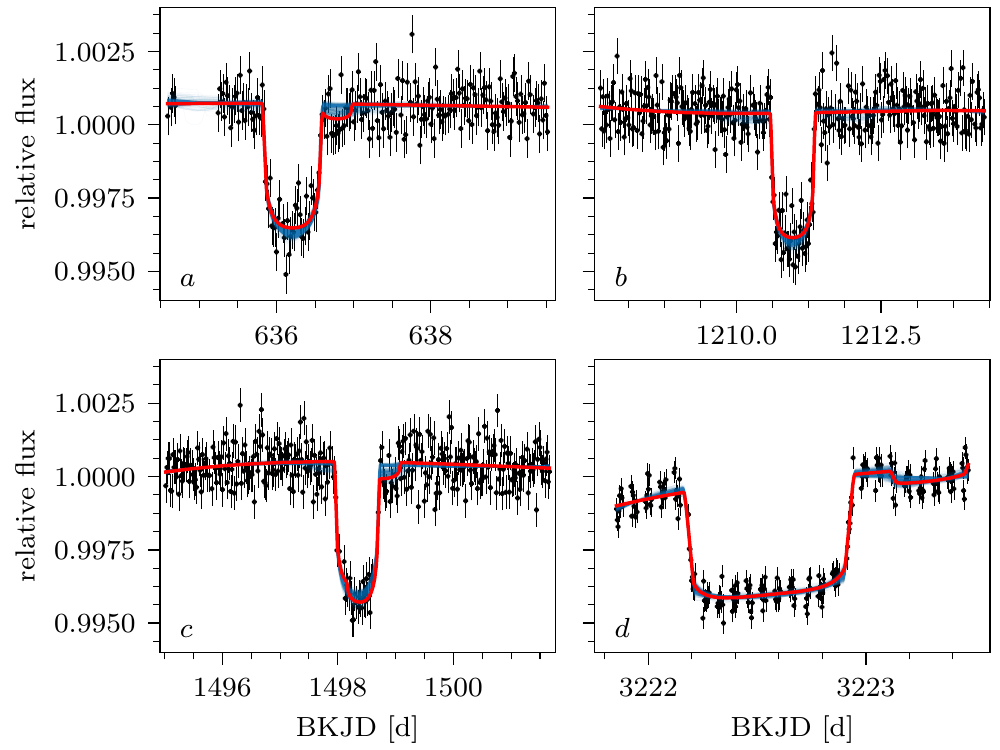}
\hfill
\includegraphics[width=0.49\textwidth]{bruno_buildin_with_inclination_one_moon_overlay_arrow_v2.pdf}
\caption{Orbital solutions for Kepler-1625\,b and its suspected exomoon based on the combined {\it Kepler} and {\it Hubble} data. ({\it a,b,c}) {\it Kepler} PDCSAP flux and ({\it d}) the quadratic detrending of the {\it Hubble} data from \citet{2018SciA....4.1784T}. The blue curves show $1\,000$ realizations of our PTMCMC fitting of a planet-moon model. Our most likely solution (red line) is very similar to the one found by \citet{2018SciA....4.1784T}, but differs significantly from the one initially found by \citet{2018AJ....155...36T}. ({\it e,f,g}) {\it Kepler} PDCSAP flux and ({\it h}) our own detrending of the {\it Hubble} light curve (in parallel to the fitting). The ingress and egress of the model moon are denoted with arrows and labels in panel {\it h} as an example.}
\label{fig:MCMC_TeacheyKipping}
\end{figure*}

\subsubsection{Mass-orbit constraints for a close-in planet}

According to \citet{2018AJ....155...36T}, the 2017 {\it Hubble} transit of Kepler-1625\,b occurred about 77.8\,min earlier than predicted, an effect that could be astrophysical in nature and is referred to as a transit timing variation (TTV). As proposed by \citet{2018AJ....155...36T}, this TTV could either be interpreted as evidence for an exomoon or it could indicate the presence of a hitherto unseen additional planet. Various planetary configurations can cause the observed TTV effect such as an inner planet or an outer planet. At this point, no stellar radial velocity measurements of Kepler-1625 exist that could be used to search for additional nontransiting planets in this system.

In the following, we focus on the possibility of an inner planet with a much smaller orbital period than Kepler-1625\,b simply because it would have interesting observational consequences. We use the approximation of \citet{2005MNRAS.359..567A} for the TTV amplitude (${\delta}t$) due to a close inner planet, which would impose a periodic variation on the position of the star, and solve their expression for the mass of the inner planet ($M_{\rm p,in}$) as a function of its orbital semimajor axis ($a_{\rm p,in}$),

\begin{equation}\label{eq:TTV}
M_{\rm p,in} = {\delta}t  \ M_\star \frac{a_{\rm p,out}}{a_{\rm p,in} \ P_{\rm p,out}} \ ,
\end{equation}

\noindent
where $a_{\rm out}=0.87$\,AU is the semimajor axis of Kepler-1625\,b. The validity of this expression is restricted to coplanar systems without significant planet-planet interaction and with $a_{\rm out}{\gg}a_{\rm in}$, so that TTVs are only caused by the reflex motion of the star around its barycenter with the inner planet.

As we show in Sect.~\ref{sec:TTVs}, the proposed inner planet could be a hot Jupiter. The transits of a Jupiter-sized planet, however, would be visible in the {\it Kepler} data. As a consequence, we can estimate the minimum orbital inclination ($i$) between Kepler-1625\,b and the suspected planet to prevent the latter from showing transits. This angle is given as per $i=\arctan(R_\star/a_{\rm p,in} )$ and we use $R_\star~=~1.793_{-0.488}^{+0.263}\,R_\odot$ \citep{2017ApJS..229...30M}.

\subsubsection{Orbital stability}

We can exclude certain masses and orbital semimajor axes for an unseen inner planet based on the criterion of mutual Hill stability. This instability region depends to some extent on the unknown mass of Kepler-1625\,b. Mass estimates can be derived from a star-planet-moon model, but these estimates are irrelevant if the observed TTVs are due to an unseen planetary perturber. Hence, we assume a nominal Jupiter mass ($M_{\rm Jup}$) for Kepler-1625\,b.

The Hill sphere of a planet with an orbital semimajor axis $a_{\rm p}$ around a star with mass $M_\star$ can be estimated as $R_{\rm H}~=~a_{\rm p}(M_{\rm p}/[3M_\star])^{1/3}$, which suggests $R_{\rm H}=125\,R_{\rm Jup}$ for Kepler-1625\,b. We calculate the Hill radius of the proposed inner planet accordingly, and identify the region in the mass-semimajor axis diagram of the inner planet that would lead to an overlap of the Hill spheres and therefore to orbital instability.

\section{Results}
\label{sec:results}

\subsection{PTMCMC sampling and $\Delta$BIC}

\begin{figure*}[t!]
\centering
\includegraphics[angle=0,width=0.49\linewidth]{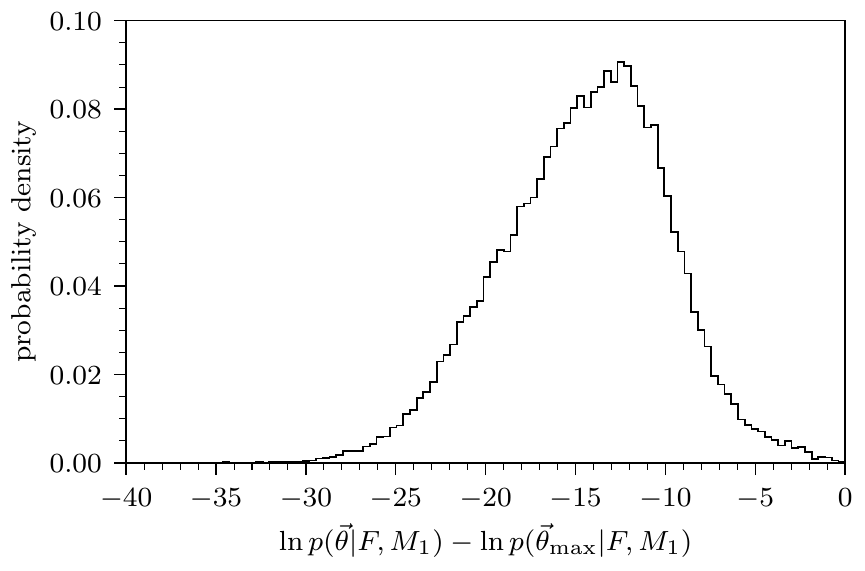}
\includegraphics[angle=0,width=0.49\linewidth]{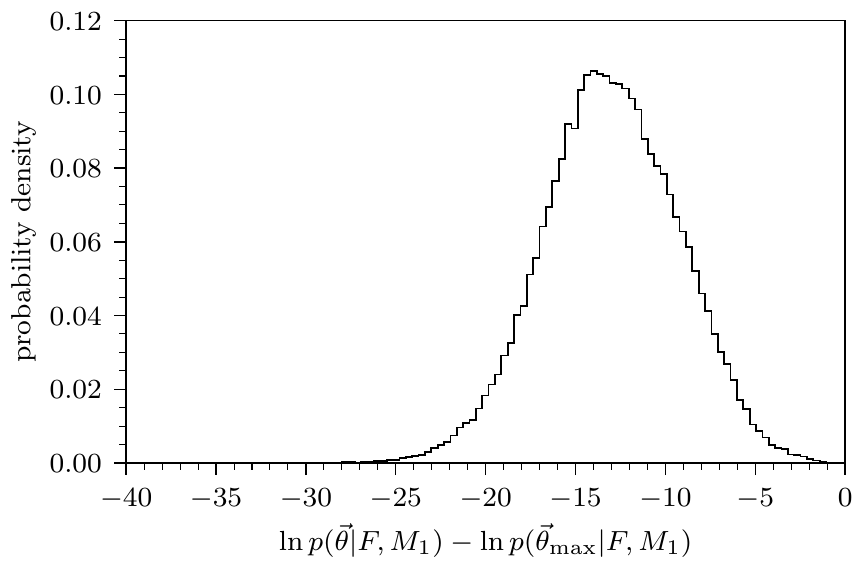}
\caption{Differential likelihood distribution between the most likely planet-moon model and the other solutions using $10^6$ steps of our PTMCMC fitting procedure. {\it Left}: Results from fitting our planet-moon transit model to the original data from \citet{2018SciA....4.1784T}. {\it Right}: Results from fitting our planet-moon transit model to our own detrending of the {\it Kepler} and WFC3 data. In both panels the most likely model is located at 0 along the abscissa by definition. In both cases the models do not converge to the best-fit solution, suggesting that the best-fit solution could in fact be an outlier.}
\label{fig:DeltaBIC}
\end{figure*}

\begin{figure*}
\centering
\includegraphics[angle=0,width=0.6\linewidth]{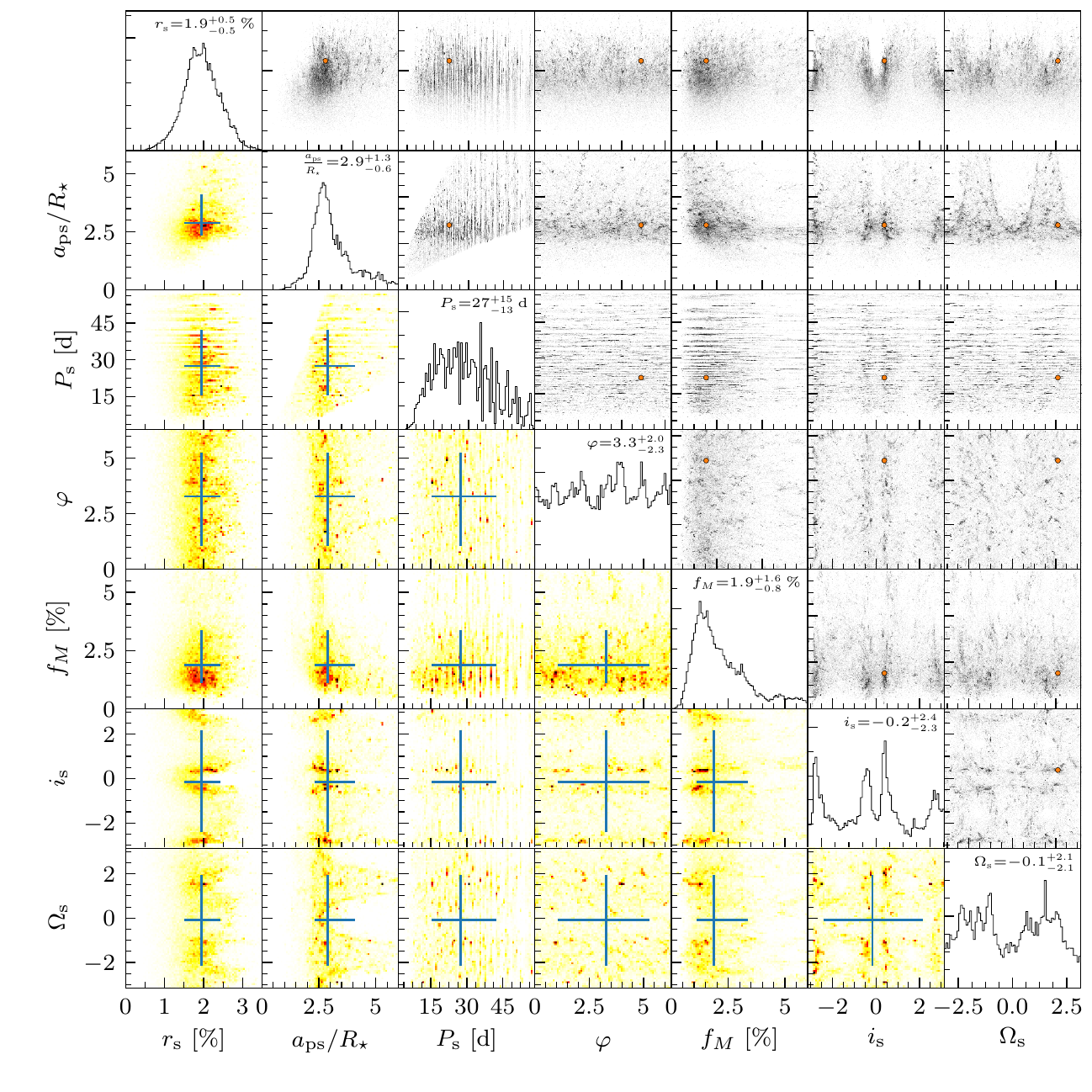}\\ \vspace*{0.15cm}
\includegraphics[angle=0,width=0.6\linewidth]{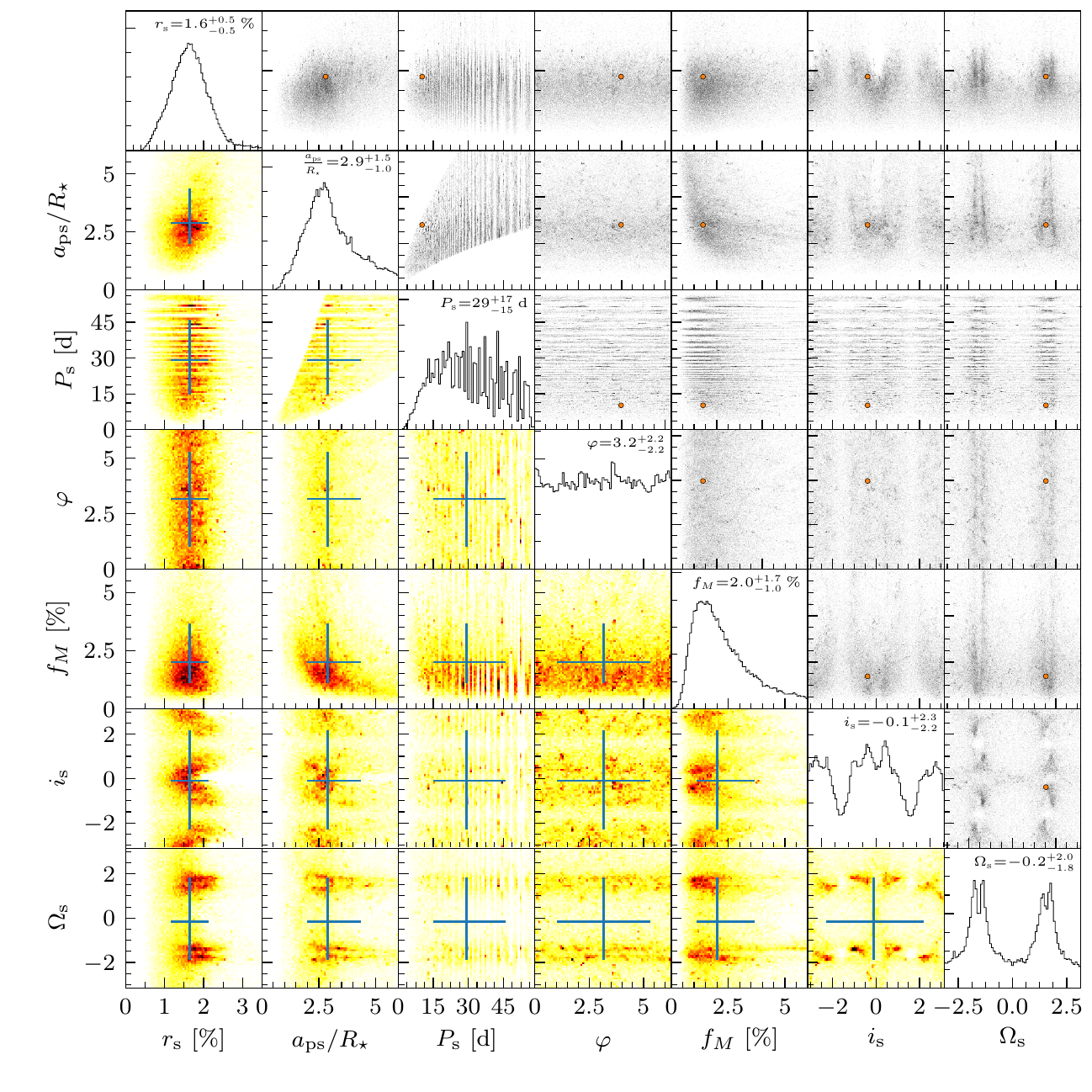}
\caption{Posterior distributions of a parallel tempering ensemble MCMC sampling of the combined {\it Kepler} and WFC3 data with our planet-moon model. {\it Top}: Results for the original data from \citet{2018SciA....4.1784T}. {\it Bottom}: Results for our own detrending of the {\it Kepler} and WFC3 data. In both figures, scatter plots are shown with black dots above the diagonal, and projected histograms are shown as colored pixels below the diagonal. The most likely parameters are denoted with an orange point in the scatter plots. Histograms of the moon-to-star radius ratio $r_{\rm s}$, scaled semimajor axis of the planet-moon system ($a_{\rm ps}/R_\star$), satellite orbital period ($P_{\rm s}$), satellite orbital phase ($\varphi$), moon-to-planet mass ratio ($f_{\rm M}$), orbital inclination of the satellite with respect to our line of sight ($i_{\rm s}$), and the orientation of the ascending node of the satellite orbit ($\Omega_{\rm s}$) are shown on the diagonal. Median values and standard deviations are indicated with error bars in the histograms.}
\label{fig:MCMCcornerplot}
\end{figure*}

Regarding the combined data set of the {\it Kepler} and {\it Hubble} data as detrended by \citet{2018SciA....4.1784T}, we find a $\Delta$BIC of $-44.5$ between the most likely planet-only and the most likely planet-moon solution. A combination of the {\it Kepler} and {\it Hubble} light curves based on our own extraction of the WFC3 data yields a $\Delta$BIC of $-31.0$. Formally speaking, both of these two values can be interpreted as strong statistical evidence for an exomoon interpretation. The two values are very different, however, which suggests that the detrending of the {\it Hubble} data has a significant effect on the exomoon interpretation. In other words, this illustrates that the systematics are not well-modeled and poorly understood.

In Fig.~\ref{fig:MCMC_TeacheyKipping}a-d, we show our results for the PTMCMC fitting of our planet-moon model to the four transits of Kepler-1625\,b including the {\it Hubble} data as extracted and detrended by \citet{2018SciA....4.1784T} using a quadratic fit. Although our most likely solution shows some resemblance to the one proposed by \citet{2018SciA....4.1784T}, we find that several aspects are different. As an example, the second {\it Kepler} transit (Fig.~\ref{fig:MCMC_TeacheyKipping}b) is fitted best without a significant photometric moon signature, that is to say, the moon does not pass in front of the stellar disk\footnote{\citet{2019arXiv190106366M} estimate that failed exomoon transits should actually be quite common for misaligned planet-moon systems, such as the one proposed by \citet{2018SciA....4.1784T}.}, whereas the corresponding best-fit model of \citet{2018SciA....4.1784T} shows a clear dip prior to the planetary transit (see their Fig.~4). What is more, most of our orbital solutions (blue lines) differ substantially from the most likely solution (red line). In other words, the orbital solutions do not converge and various planet-moon orbital configuration are compatible with the data, though with lower likelihood.

In Fig.~\ref{fig:MCMC_TeacheyKipping}e-h, we illustrate our results for the PTMCMC fitting of our planet-moon model to the four transits of Kepler-1625\,b including our own extraction and detrending of the {\it Hubble} transit. Again, the orbital solutions (blue lines) do not converge. A comparison of panels d and h shows that the different extraction and detrending methods do have a significant effect on the individual flux measurements, in line with the findings of \citet{2018A&A...617A..49R}. Although the time of the proposed exomoon transit is roughly the same in both panels, we find that the best-fit solution for the data detrended with our own reduction procedure does not contain the moon egress (panel h), whereas the best-fit solution of the data detrended by \citet{2018SciA....4.1784T} does contain the moon egress (panel d). A similar fragility of this particular moon egress has been noted by \citet{2018SciA....4.1784T} as they explored different detrending functions (see their Fig.~3).

Our Fig.~\ref{fig:DeltaBIC} illustrates the distribution of the differential likelihood for the planet-moon model between the most likely model parameter set ($\vv{\theta}_{\rm max}$) and the parameter sets ($\vv{\theta}'$) found after five million steps of our PTMCMC fitting procedure, $p(\vv{\theta}'|F,\mathcal{M}_1) - p(\vv{\theta}_{\rm max}|F,\mathcal{M}_1)$. For the combined {\it Kepler} and {\it Hubble} data detrended by \citet{2018SciA....4.1784T} (left panel) and for our own {\it Hubble} data extraction and detrending (right panel), we find that most model solutions cluster around a differential likelihood that is very different from the most likely solution, suggesting that the most likely model is, in some sense, a statistical outlier. We initially detected this feature after approximately the first one hundred thousand PTMCMC fits. Hence, we increased the number of PTMCMC samplings to half a million and finally to five million to make sure that we sample any potentially narrow peaks of the likelihood function near the best-fit model at $p(\vv{\theta}'|F,\mathcal{M}_1) - p(\vv{\theta}_{\rm max}|F,\mathcal{M}_1)=0$ with sufficient accuracy. We find, however, that this behavior of the differential likelihood distribution clustering far from the best-fit solution persists, irrespective of the available computing power devoted to the sampling.

\begin{table}
\caption{Results of our PTMCMC fitting procedure to the combined {\it Kepler} and {\it Hubble} data. The {\it Hubble} data was either based on the photometry extracted by \citet[][TK18b, central column]{2018SciA....4.1784T} or based on our own extraction (right column).}
\def\arraystretch{2}
\label{tab:MCMC}
\small 
\centering
\begin{tabular}{c c c}
\hline\hline
& TK18b {\it HST} photometry & Our {\it HST} photometry \\
\hline
$r_{\rm s}$\,[\%] & $1.9_{-0.5}^{+0.5}$ & $1.6_{-0.5}^{+0.5}$ \\
$a_{\rm ps}$\,[$R_{\star}$] & $2.9_{-0.6}^{+1.3}$ & $2.9_{-1.0}^{+1.5}$\\
$P_{\rm s}$\,[d] & $27_{-13}^{+15}$ & $29_{-15}^{+17}$ \\
$\varphi$\,[rad] & $3.3_{-2.3}^{+2.0}$ & $3.2_{-2.2}^{+2.2}$ \\
$f_{\rm M}$\,[\%] & $1.9_{-0.8}^{+1.6}$ & $2.0_{-1.0}^{+1.7}$ \\
$i_{\rm s}$\,[rad] & $-0.2_{-2.3}^{+2.4}$ & $-0.1_{-2.2}^{+2.3}$\\
$\Omega_{\rm s}$\,[rad] & $-0.1_{-2.1}^{+2.1}$ & $-0.2_{-1.8}^{+2.0}$\\
\hline
\end{tabular}
\tablefoot{Figure~\ref{fig:MCMCcornerplot} illustrates quite clearly that the posterior distributions are not normally distributed and often not even representative of skewed normal distributions. The confidence intervals stated in this table have thus to be taken with care.}\\
\end{table}

Figure~\ref{fig:MCMCcornerplot} shows the posterior distributions of the moon parameters of our planet-moon model. The top panel refers to our PTMCMC fitting of the combined {\it Kepler} and {\it Hubble} data ({\it Hubble} data as detrended and published by \citealt{2018SciA....4.1784T}), and the bottom panel shows our PTMCMC fitting of the {\it Kepler} data combined with our own extraction and detrending of the {\it Hubble} light curve. The respective median values and standard deviations are noted in the upper right corners of each subpanel and summarized in Table~\ref{tab:MCMC}.

A comparison between the upper and lower corner plots in Fig.~\ref{fig:MCMCcornerplot} reveals that the different detrending and fitting techniques have a significant effect on the resulting posterior distributions, in particular for $i_{\rm s}$ and $\Omega_{\rm s}$, the two angles that parameterize the orientation of the moon orbit. At the same time, however, the most likely values (red dots above the plot diagonal) and median values (blue crosses below the plot diagonal) of the seven parameters shown are well within the $1\,\sigma$ tolerance.

The following features can be observed in both panels of Fig.~\ref{fig:MCMCcornerplot}. The moon-to-star radius ratio (Col. 1, leftmost) shows an approximately normal distribution, whereas the scaled planet-moon orbital semimajor axis (Col. 2) shows a more complicated, skewed distribution. The solutions for the orbital period of the exomoon candidate (Col. 3) show a comb-like structure owing to the discrete number of completed moon orbits that would fit a given value of the moon's initial orbital phase (Col. 4), which is essentially unconstrained. The moon-to-planet mass ratio (Col. 5) then shows a skewed normal distribution with a tail of large moon masses. Our results for the inclination $i_{\rm s}$ between the satellite orbit (around the planet) and the line of sight, and for the longitude of the ascending node of the moon orbit are shown in Cols. 6 and 7. The preference of $i_{\rm s}$ being either near 0 or near $\pm\pi$ (the latter is equivalent to a near-coplanar retrograde moon orbit) illustrates the well-known degeneracy of the prograde/retrograde solutions available from light curve analyses \citep{2014ApJ...791L..26L,2014ApJ...796L...1H}.

\subsection{Transit timing variations}
\label{sec:TTVs}

\begin{figure}[t!]
\centering
\includegraphics[angle= 0, width=1\linewidth]{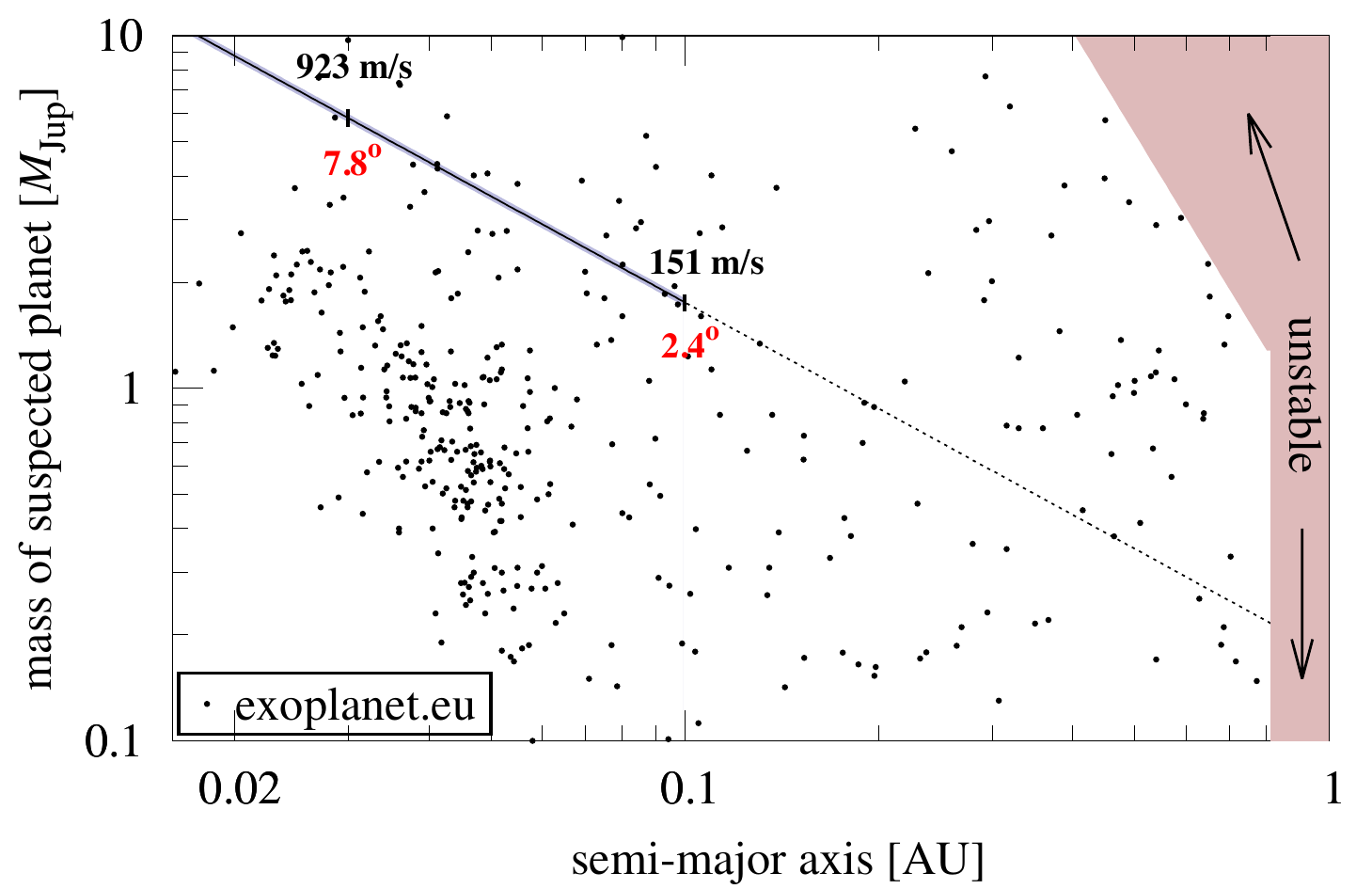}
\caption{Mass estimate for the potential inner planet around Kepler-1625 based on the observed TTV of 73.728\,min. The thin pale blue fan around the solid curve shows the $1\,\sigma$ tolerance fan of $\pm2.016$\,min. Values for semimajor axes $>0.1$\,AU are poor approximations and thus shown with a dashed line. Black points show masses and semimajor axes of all planets from \href{exoplanet.eu}{exoplanet.eu} (as of 26 October 2018) around stars with masses between $0.75\,M_\odot$ and $1.25\,M_\odot$. A conservative estimate of a dynamically unstable region for the suspected inner planet, where its Hill sphere would touch the Hill sphere of Kepler-1625\,b with an assumed mass of $1\,M_{\rm Jup}$, is shaded in pale red. RV amplitudes and minimum orbital inclination with respect to Kepler-1625\,b are noted along the curve for the planetary mass estimate.}
\label{fig:TTV_Mp}
\end{figure}

Next we consider the possibility of the transits being caused by a planet only. Neglecting the {\it Hubble} transit, our PTMCMC sampling of the three {\it Kepler} transits with our planet-only transit model gives an orbital period of $P=287.3776\,\pm~0.0024$\,d and an initial transit midpoint at $t_0=61.4528\,\pm~0.0092$\,d in units of the Barycentric Kepler Julian Day (BKJD), which is equal to ${\rm BJD} - 2,454,833.0$\,d. The resulting transit time of the 2017 {\it Hubble} transit is $3222.6059\,\pm~0.0182$\,d.

Our planet-only model for the 2017 {\it Hubble} transit gives a transit midpoint at $3222.5547\,\pm0.0014$\,d, which is $73.728 (\pm2.016)$\,min earlier than the predicted transit midpoint. This is in agreement with the measurements of \citet{2018SciA....4.1784T}, who found that the {\it Hubble} transit occurred 77.8\,min earlier than predicted. This observed early transit of Kepler-1625\,b has a formal $\sim3\,\sigma$ significance. We note, however, that this $3\,\sigma$ deviation is mostly dictated by the first transit observed with {\it Kepler} \citep[see Fig.~S12 in][]{2018AJ....155...36T}. We also note that this transit was preceded by a $\sim1$\,d observational gap in the light curve, about 0.5\,d prior to the transit, which might affect the local detrending of the data and the determination of the transit mid-point of a planet-only model. Moreover, with most of the TTV effect being due to the large deviation from the linear ephemeris of the first transit, stellar (or any other systematic) variability could have a large (but unknown) effect on the error bars that go into the calculations.

In Fig.~\ref{fig:TTV_Mp} we show the mass of an unseen inner planet that is required to cause the observed $73.7$\,min TTV amplitude from our PTMCMC fit as a function of its unknown orbital semimajor axis. The mass drops from $5.8\,M_{\rm Jup}$ at 0.03\,AU to $1.8\,M_{\rm Jup}$ at 0.1\,AU. Values beyond 0.1\,AU cannot be assumed to fulfill the approximations made for Eq.~\eqref{eq:TTV} and are therefore shown with a dashed line. The actual TTV amplitude of Kepler-1625\,b could even be higher than the $\sim73$\,min that we determined for the {\it Hubble} transit, and thus the mass estimates shown for a possible unknown inner planet serve as lower boundaries.

The resulting radial velocity amplitudes of the star of $923\,{\rm m\,s}^{-1}$ (at 0.03\,AU) and $151\,{\rm m\,s}^{-1}$ (at 0.1\,AU), respectively, are indicated along the curve. Even if the approximations for a coplanar, close-in planet were not entirely fulfilled, our results suggest that RV observations of Kepler-1625 with a high-resolution spectrograph attached to a very large (8\,m class) ground-based telescope could potentially reveal an unseen planet causing the observed TTV of Kepler-1625\,b. Also shown along the curve in Fig.~\ref{fig:TTV_Mp} are the respective minimum orbital inclinations (rounded mean values shown) between Kepler-1625\,b and the suspected close-in planet required to prevent Kepler-1625\,b from transiting the star. The exact values are $i=7.8_{-2.0}^{+1.1}$ degrees at 0.03\,AU and $i=2.4_{-0.6}^{+0.3}$ degrees at 0.1\,AU.

The pale red shaded region is excluded from a dynamical point of view since this is where the planetary Hill spheres would overlap. The extent of this region is a conservative estimate because it assumes a mass of $1\,M_{\rm Jup}$ for Kepler-1625\,b and neglects any chaotic effects induced by additional planets in the system or planet-planet cross tides etc. The true range of unstable orbits is probably larger. The black dots show all available exoplanet masses and semimajor axes from the Exoplanet Encyclopaedia, which illustrates that the suspected planet could be more massive than most of the known hot Jupiters.

\section{Conclusions}

With a $\Delta$BIC of $-44.5$ \citep[using published {\it Hubble} data of][]{2018SciA....4.1784T} or $-31.0$ (using our own {\it Hubble} extraction and detrending) between the most likely planet-only model and the most likely planet-moon model, we find strong statistical evidence for a roughly Neptune-sized exomoon. In both cases of the data detrending, the most likely orbital solution of the planet-moon system, however, is very different from most of the other orbital realizations of our PTMCMC modeling and the most likely solutions do not seem to converge. In other words, the most likely solution appears to be an outlier in the distribution of possible solutions and small changes to the data can have great effects on the most likely orbital solution found for the planet-moon system. As an example, we find that the two different detrending methods that we explored produce different interpretations of the transit observed with {\it Hubble}: in one case our PTMCMC sampling finds the egress of the moon in the light curve, in the other case it does not (Fig.~\ref{fig:MCMC_TeacheyKipping}).

Moreover, the likelihood of this best-fit orbital solution is very different from the likelihoods of most other solutions from our PTMCMC modeling. We tested both a standard MCMC sampling and a parallel-tempering MCMC \citep{2013PASP..125..306F}; the latter is supposed to explore both the parameter space at large and the tight peaks of the likelihood function in detail. Our finding of the nonconvergence could imply that the likelihood function that best describes the data is non-Gaussian. Alternatively, with the BIC being an asymptotic criterion that requires a large sample size by definition \citep{2012ApJ...754..136S}, our findings suggest that the available data volume is simply too small for the BIC to be formally applicable. We conclude that the ${\Delta}$BIC is an unreliable metric for an exomoon detection for this data set of only four transits and possibly for other data sets of {\it Kepler} as well.

One solution to evaluating whether the BIC or an alternative information criterion such as the Akaike information criterion \citep[AIC;][]{1974ITAC...19..716A} or the deviance information criterion \citep[DIC;][]{Spiegelhalter2002} is more suitable for assessing the likelihoods of a planet-only model and of a planet-moon model could be injection-retrieval experiments of synthetic transits \citep{2016A&A...591A..67H,2018A&A...617A..49R}. Such an analysis, however, goes beyond the scope of this paper.

We also observe the TTV effect discovered by \citet{2018SciA....4.1784T}. If the early arrival of Kepler-1625\,b for its late-2017 transit was caused by an inner planet rather than by an exomoon, then the planet would be a super-Jovian mass hot Jupiter, the exact mass limit depending on the assumed orbital semimajor axis. For example, the resulting stellar radial velocity amplitude would be about $900\,{\rm m\,s}^{-1}$ for a $5.8\,M_{\rm Jup}$ planet at 0.03\,AU and about $150\,{\rm m\,s}^{-1}$ for a $1.8\,M_{\rm Jup}$ planet at 0.1\,AU. From the absence of a transit signature of this hypothetical planet in the four years of {\it Kepler} data, we conclude that it would need to have an orbital inclination of at least $i=7.8_{-2.0}^{+1.1}$ (if it were at 0.03\,AU) or $i=2.4_{-0.6}^{+0.3}$ degrees (if it were at 0.1\,AU). If its inclination is not close to $90^\circ$, at which point its effect on the stellar RV amplitude would vanish, then the hypothesis of an unseen inner planet causing the Kepler-1625 TTV could be observationally testable.

Ground-based photometric observations are hardly practicable to answer the question of this exomoon candidate because continuous in- and near-transit monitoring of the target is required over at least two days. Current and near-future space-based exoplanet missions, on the other hand, will likely not be able to deliver the signal-to-noise ratios required to validate or reject the exomoon hypothesis. With a {\it Gaia} G-band magnitude of $m_{\rm G}=15.76$ \citep{2016A&A...595A...1G,2018A&A...616A...1G} the star is rather faint in the visible regime of the electromagnetic spectrum and the possible moon transits are therefore beyond the sensitivity limits of the {\it TESS}, {\it CHEOPS}, and {\it PLATO} missions. {\it 2MASS} observations suggest that Kepler-1625 is somewhat brighter in the near-infrared \citep{2003yCat.2246....0C}, such that the {\it James Webb Space Telescope} (launch currently scheduled for early 2021) should be able to detect the transit of the proposed Neptune-sized moon, for example via photometric time series obtained with the NIRCam imaging instrument.

All things combined, the fragility of the proposed photometric exomoon signature with respect to the detrending methods, the unknown systematics in both the {\it Kepler} and the {\it Hubble} data, the absence of a proper assessment of the stellar variability of Kepler-1625, the faintness of the star (and the resulting photometric noise floor), the previously stated coincidence of the proposed moon's properties with those of false positives \citep{2018A&A...617A..49R}, the existence of at least one plausible alternative explanation for the observed TTV effect of Kepler-1625\,b, and the serious doubts that we have about the $\Delta$BIC as a reliable metric at least for this particular data set lead us to conclude that the proposed moon around Kepler-1625\,b might not be real. We find that the exomoon hypothesis heavily relies on a chain of delicate assumptions, all of which need to be further investigated.

A similar point was raised by \citet{2018SciA....4.1784T}, and our analysis is an independent attempt to shed some light on the ``unknown unknowns'' referred to by the authors. For the time being, we take the position that the first exomoon has yet to be detected as the likelihood of an exomoon around Kepler-1625\,b cannot be assessed with the methods used and data currently available.

\begin{acknowledgements}
The authors thank Kevin Stevenson, Hannah Wakeford and Megan Sosey for their help with the data analysis, and Nikole Lewis for the feedback on the manuscript. The authors would also like to thank the referee for a challenging and constructive report. This work was supported in part by the German space agency (Deutsches Zentrum f\"ur Luft- und Raumfahrt) under PLATO Data Center grant 50OO1501. This work made use of NASA's ADS Bibliographic Services and of the Exoplanet Encyclopaedia (\href{http://exoplanet.eu}{http://exoplanet.eu}). RH wrote the manuscript, proposed Figs.~\ref{fig:MCMC_TeacheyKipping} - \ref{fig:MCMCcornerplot}, generated Fig.~\ref{fig:TTV_Mp}, and guided the work. KR derived the star-planet-moon orbital simulations and the respective statistics and generated Figs.~\ref{fig:MCMC_TeacheyKipping} - \ref{fig:MCMCcornerplot}. GB performed the light curve extraction from the WFC3 {\it Hubble} data and generated Fig.~\ref{fig:HST}. All authors contributed equally to the interpretation of the data.
\end{acknowledgements}

\bibliographystyle{aa}
\bibliography{ms}





\end{document}